Nanoassembly of Polydisperse Photonic Crystals based on Binary and Ternary Polymer Opal Alloys

*Qibin Zhao, Chris E. Finlayson, Christian Schafer, Peter Spahn, Markus Gallei, Lars Herrmann, Andrei Petukhov, and Jeremy J. Baumberg\**

Dr. Qibin Zhao, Dr. Lars Herrmann, Prof. Jeremy J. Baumberg
Cavendish Laboratory, University of Cambridge, Cambridge CB3 0HE, United Kingdom
E-mail: jjb12@cam.ac.uk

Dr. Chris E. Finlayson
Department of Physics, Prifysgol Aberystwyth University, Aberystwyth, Wales SY23 3BZ, United Kingdom

Dr. Christian Schafer, Dr. Peter Spahn, Dr. Markus Gallei
Ernst-Berl Institute for Chemical Engineering and Macromolecular Science, Darmstadt University of Technology, Darmstadt D-64287, Germany

Dr. Andrei Petukhov
Van't Hoff Laboratory for Physical and Colloid Chemistry, Debye Research Institute, Utrecht University, Utrecht 3584 CH, The Netherlands

Since the first discovery of binary opals by Sanders and Murray several decades ago in Brazilian gem opals,[1] colloidal alloys have attracted increasing attention due to their possible applications in various fields.[2] A notable feature of such alloys is the ability to combine two or more distinct components to produce binary, ternary, or even higher order opals with more complex periodicities and optical properties.[3] Moreover, the possibility of generating complete bandgap photonic crystals based on diamond or pyrochlore structures has also been shown.[4]

Research in this field has progressed significantly following successes in single component colloidal photonic crystal assembly.[5] Indeed, most of the structural fabrication methods in colloidal alloys are derived from single component colloidal assembly. Depending on whether different components are assembled at the same time in the same mixture, the methods may be categorized into either layer-by-layer or one-stage approaches.[6] Different components are assembled sequentially in the former approach,[7] whereas in the one-stage approach they are mixed together and self-assemble simultaneously into alloys.[8] These methods can also be distinguished by whether the process is self-driven or induced by external fields.[9, 10] More specific structures can be made by combining self-assembly with the use of controlling external fields.[11]

The size ratio between different components is a critical parameter in the generation of colloidal alloys. Thus far, systems with a diameter ratio between the small and large components ($D_{S/L}$) of less than ~0.3 are considered to be energetically unstable for self-assembly, and generally $D_{S/L} > 0.6$ is very rare for colloidal alloys, regardless of the method used, so that $D_{S/L}$ of ~0.5 is required for any degree of success in alloying.[12] When $D_{S/L}$ is large (i.e. close to 1), it can be compared with the effects of polydispersity in single-component colloidal systems, which strongly inhibits crystallization.[9, 13]

Here, we report binary and ternary polymer opal (PO) alloys with a high degree of structural ordering, made by a bending induced oscillatory shearing (BIOS) method.[14] Oscillatory shearing methods including BIOS have been previously shown to be highly effective for inducing crystallization in single component POs on large scales, indicating a wide range of promising applications in next generation bulk-scale photonic structures, coatings, fibres, and sensors.[15] However, the success reported here in making ordered binary and ternary PO alloys demonstrates the exceptional ability of BIOS for inducing order in solvent-free viscoelastic systems of particles. The components used in our experiments are hard-core/soft-shell spheres of three distinct outer radii.[16, 17] These radii of 92nm, 110nm, and 130nm (with polydispersity for each size of 2-4%) therefore access diameter ratios from $D_{S/L} = 0.71$-$0.85$, far beyond the limits for previous colloidal alloys. Since the single-size PO components used give photonic stop bands in the blue, green, and red parts of the spectrum, the three components are denoted as B, G, and R respectively going from small to large (**Figure 1**). Binary PO alloys comprised of two components, with a ratio between the number of spheres ($N_{S/L}$) = 1:1, and ternary PO alloys comprised of all three size components with different number ratios $N_{B/G/R}$ = 3:2:1, 1:1:1, and 1:2:3 are fabricated. As shown below, a clear layered structure parallel to the sample surface is observed in both binary and ternary PO alloys after BIOS processing, and a hexagonal pattern is found to exist in the in-plane arrangement of the layers with preferred close-packed directions parallel to the shearing direction.

To our knowledge, these results are the first to show such strong structural color in binary and ternary alloys with high size ratio between component spheres, and also the first example of a viscoelastic opal alloy. Different colors may then be generated by mixing different sized spheres together, using a completely alternative mechanism to the blending of pigments in subtractive color. Previous manifestations of binary/ternary colloidal alloys are rarely more than 20 layers, however the thickness of the PO alloy films studied here is 80μm or nearly 1000 layers, and the nature of processing is readily compatible with bulk-scale sample production. We can also make comparisons with athermal granular systems, in which close-packing ('jamming') emerges near a critical volume fraction, below which the material is a fluid whose viscosity increases rapidly.[18] However, crystallization (specifically crystalline ordering) is prevented by a critical level of polydispersity.[19, 20] These prior studies reinforce the special nature of the polymer opal system here. Flow ordered colloidal suspensions are solvent mediated and totally different to the kinetically-dominated high-viscosity contact regime explored here. Typical colloidal suspensions are also entropy driven and the equilibrium structures determined by interparticle collisions, dispersive/diffusive forces, and electrostatic forces. By contrast, the equilibrium in polymer opals is mainly associated with the accumulation and release of strain energy from external forces such as shearing and stretching.[21] We anticipate such developments in assembling functional nanostructures within viscoelastic media (and using the BIOS process in particular) have general applicability to fabricate many other composite optical materials yielding greatly reduced requirements for monodispersity.

The polymer opal materials studied here are based on ensembles of core-interlayer-shell particles, as illustrated in Figure 1. These are synthesized using a multi-stage emulsion polymerization process, as previously reported.[16, 22] The particle precursors range in size from 180 to 260 nm in outer diameter, and consist of a hard heavily-cross-linked polystyrene (PS) core, coated with a polymethacrylate-based grafting layer, and a soft polyethylacrylate (PEA) outer-shell. The core-shell design of the particles means that even at the point of melting, the polymer opals do not contain a discrete fluid phase. The net refractive index contrast between core and shell materials is $\Delta n \approx 0.11$, and the volume fraction of cores $\approx 55\%$.

Polymer opal alloy thin-film samples are then made following the same procedures as previously reported for single component POs[23] (see also Figure 1d). Spheres of different sizes with designated number ratios are mixed in a polymer extruder and squeezed into ribbons through a rectangular die. The ribbons from extrusion are then rolled into thin films with a standard thickness of $\sim 80\mu m$ and laminated between two PET sheets of high Young's modulus. Following on from this standard fabrication protocol, the BIOS process[14] is then applied to the sandwich structure immediately after the continuous rolling lamination, and an ordered PO alloy layer is obtained by repetition of oscillatory shearing, as illustrated in Figures 1c,d. Effectively the BIOS processing is achieved by mechanically cycling (oscillating) the sandwich structure under tension around a fixed cylindrical surface (here of diameter 12 mm) at a stabilized temperature of 100°C. Since the PET encapsulant film is far more rigid than the PO, the PET-PO-PET laminate constitutes a Timoshenko sandwich beam.[14] Hence, by bending the laminate around a cylinder, strong shear is created inside the PO purely parallel to the surface, thus generating strains of magnitude up to 300%.[14]

The experimental methods associated with subsequent optical microscopy, small angle X-ray scattering (SAXS), optical reflectometry, scanning electron microscopy (SEM) and focussed ion beam SEM (FIB-SEM) are detailed in the *Experimental Methods* section.

After the rolling lamination step, the PO alloy films are generally milky-white in color (**Figure 2a**), with very slight hues due to size-dependent incoherent multiple scattering from the spheres,[24] implying that the configurational arrangement of the spheres is essentially random. No resonant peaks emerge in the measured reflection spectra (Figure 2b), and SAXS patterns at normal incidence to the sample surface show the typical diffraction rings expected for an amorphous structure (as will be described below). Very weak $(10)$ and $(\bar{1}0)$ spots are found in the lateral direction of the patterns, which indicates a slightly preferred orientation of the spheres in the rolling direction. Ordered structures start to form upon application of the BIOS process. Resonant Bragg peaks appear as signatures for the formation of layered structures, and grow in intensity with increased amounts of oscillatory shearing (Figure 2c), which now give structural color to the PO alloys. In a similar fashion to single component POs, growth of the ordered structure reaches equilibrium after ~40 passes.[14]

Images of the fabricated PO alloys after BIOS are also shown in Figure 2. As readily apparent to the naked eye, the hue/saturation of the colors of the PO alloys varies with the number ratios between different components. This is confirmed by the measured optical reflectivities at normal incidence. We also note the differences in Bragg peak amplitude for the single component blue, green and red POs (top), which we attribute to minor variations in the rheological properties of the spheres arising from the emulsion polymerization process that do not greatly affect our conclusions below. The Bragg peaks of PO alloys have lower amplitude than those from the single component samples. Surprisingly, however, the spectral positions of the peaks from the PO alloys are not simply the mean of their sub-component peak positions ($\lambda_B$, $\lambda_G$, $\lambda_R$). We further analyse the measured spectra by extracting the spectral positions and amplitudes of the resonant peaks, plotting the peak position of the PO alloys on a ternary phase diagram (**Figure 3a,b**). The peaks of the alloys are always shifted to longer wavelengths than the corresponding weighted average. For instance, the peak position of the PO alloy with ratio B:G = 1:1 is 637nm, 15nm longer than the weighted average, $\lambda_p = \lambda_B f_B + \lambda_G f_G = 622$nm, for fill fractions $f_i$. This has to be understood in terms of the dependence of the resonant peaks observed due to the ordered layers stacked in the depth direction, according to the Bragg diffraction equation

$$\lambda_p = 2d_{hkl}(n_{\text{eff}}^2 - \sin^2\theta)^{1/2} \tag{1}$$

(where $d_{hkl}$ is the layer spacing, $n_{\text{eff}}$ is the effective refractive index, and $\theta$ is the incident angle with respect to the normal of the sample surface). This implies that in alloys the layer spacing is more strongly influenced by the larger spheres in the alloys instead of being a mean average of all the spheres. As we discuss below, the larger spheres alloyed into each layer keep apart the neighbouring layers, but as they redistribute to avoid being exactly on top of other large spheres, the spacing adapts accordingly.

To quantify optically how well ordered the alloys are we study the resonant peak amplitude, normalized to the amplitude of their largest component (so for the case where B:G:R = 3:2:1 the amplitude is normalized with that of the single component red PO). We then apply the concept of polydispersity index (PDI) in order to quantify the magnitude of bimodal or trimodal particle size variation of spheres in the alloys

$$\text{PDI} = \sigma/D_0 \tag{2}$$

where $\sigma$ is the standard deviation and $D_0$ is the mean diameter of the spheres. Empirically, we find that there is a linear correlation between the normalized amplitude of the PO alloys with polydispersity (Figure 3c). The reflection amplitude drops with increasing PDI, and extrapolates to predict that no resonant peak exists for PDI > 20%, thus setting an upper tolerance limit of particle polydispersity for generating ordered PO alloys via the BIOS method. This is a four-fold improvement compared to the 5% polydispersity found to destroy all ordering in colloidal suspensions.[20]

Whilst the peak position in the optical spectra depends only on layer spacing, the amplitude of the peaks depends on both the in-plane order of the layers and the interlayer ordering. In-plane order determines the reflection from each layer, while interlayer order determines the variation of effective refractive index $n_{eff}$ in the depth direction, and thus the effective refractive index contrast $\Delta n_{eff}$. By using FIB-SEM,[25] we get clear images of the cross sections of PO alloys, in comparison with green single component POs (**Figure 4**, also Supporting Information Figures S1 and S2). The positions of spheres in each image are extracted, allowing an estimation of $n_{eff}$ ($z$) along the thickness direction, taking refractive indices for the polymeric components of $n_{PS} = 1.59$ and $n_{PEA} = 1.48$. Representative images from depths of 4–8 μm for a binary PO alloy of G:R = 1:1 and a ternary PO alloy of B:G:R = 3:2:1 are shown in Figures 4b & 4c. Clear lines of spheres are visible in single component and binary PO alloys of G:R = 1:1 deep inside the samples. Linear alignment of the spheres is also observed near the surface of ternary B:G:R = 3:2:1 alloys, however the signatures of alignment become less obvious going deeper into the sample. The layer stacking structure is well illustrated from the $n_{eff}$ spatial distributions, which show regular fluctuations (insets). Fast Fourier Transforming (FFT) the $n_{eff}$ ($z$) distributions gives clear evidence of a sharp strong central frequency in the spatial variation of the spheres in the thickness direction (Figure 3d). The central spatial frequencies, as well as $\Delta n_{eff}$ between the layers at different depths of the samples, are extracted from the FFT results for the complete set of samples (Supporting Information Figure S3). In a similar fashion to single component POs, the spatial frequency of the layers increases exponentially with increasing depth in PO alloys. The reasons for the exponential trend are due to the visco-elastic nature of the material, as discussed in our recent study of the BIOS process.[14] The resonant peaks in the optical reflectivity arise from regions near the sample surface (typically depths from 4 − 12μm) rather than from all the layers, with a spectral shape determined by this exponential spacing dependence from the surface. From Figure 4 we find that the spatial frequency of the ternary PO alloy of B:G:R = 3:2:1 is higher than that of single component green PO, while that of binary PO alloy of G:R = 1:1 is lower than that of green PO, consistent with the optical measurements. In general, $\Delta n_{eff}$ (G) > $\Delta n_{eff}$ (G:R=1:1) > $\Delta n_{eff}$ (B:G:R = 3:2:1), which indicates the increasingly strong effects of spheres penetrating into neighboring layers, induced by the increasing polydispersity. As a result, the successful accommodation of large polydispersity in the layer stacking is accompanied by a decrease in the spectral intensity.

Also surprising is the effective development of hexagonal packing of the spheres within the layers after BIOS processing the binary and ternary PO alloys (**Figure 5**). The SAXS patterns are measured at normal incidence to the surface and the samples are aligned with the shearing direction along the horizontal, allowing clear structural information from the layers to be obtained. Before shearing, scattering rings dominate the SAXS patterns, as is typical for amorphous structures.[26] Hexagonal lattice coordinates are used for indexing the reciprocal lattice of the samples (Figure 5b). After 40 passes of BIOS, clear diffraction spots in a hexagonal pattern appear in the diffuse diffraction rings. This indicates that although the systems are comprised of bimodal or trimodal components of different size, somewhat ordered structures are able to form with oscillatory shearing, and the spheres are arranged in similar fashion to the single component POs. The patterns show the same preferred orientation of the spheres, with the most closely packed (mcp) direction of the spheres being along the shearing direction. The results are analyzed by dividing the diffraction patterns into sectors, each sector containing one spot, and fitting the spots to a 2D Gaussian lineshape. Our analysis (Figure S4) shows that the integrated intensity, as well as the amplitude of the spots, increases drastically after BIOS, accompanied by a strong decrease in the FWHM along both the long axis $\Lambda r$ of the spots and the short axis $\Lambda s$. In general, the radial distance from the (10) spot to the origin increases slightly after BIOS, while the distance from the $(1\bar{1})$ spot to the origin decreases, which implies that sphere separations in the shearing direction are becoming more compact, whilst they move further apart in the direction perpendicular to the shearing direction. It is also interesting to note that the strength of spot (10) is generally higher than that of $(1\bar{1})$ spot for all the samples before shearing, however this asymmetry is removed after shearing. We argue that this is due to relative spatial offsets between layers, as described recently for single component POs.[14, 23] Beyond this, the SAXS intensities and other diffraction features from the different PO alloys cannot readily be directly compared since the form factors (which depend upon the sizes of the component particles) will be different in each case.

Finally, in order to determine whether other 3D periodic structures exist in PO alloys, such as the quasi-(200) layers found in single component POs, specular reflection measurements are performed with optical goniometry (Supporting Information Figure S5). No convincing signatures of repulsive modes due to other periodicities are observed. All of the available characterization evidence presented here indicates that crystallization of polydisperse nano-ensembles by these methods forms genuine alloy structures with random size-occupancy of each lattice site, and does not involve any separation of sizes, or site correlations.

In conclusion, ordered binary and ternary polymer opal alloys comprised of spheres with size ratio $D_{S/L}$ = 0.71-0.85 are formed using a bending induced oscillatory shearing (BIOS) method. Layered structures are observed to form parallel to the sample surface, with a spacing which shows an exponential variation with depth, and dominated (but not exclusively set by) the largest sphere. We thus obtain a different way to tune the film iridescence spectrum, by mixing R,G,B primary spheres together in alloy POs. Spheres inside the layers show a dominant hexagonal packing, with the most closely packed direction parallel to the shearing direction. We demonstrate the ability of the BIOS method to induce crystallization in diverse polydisperse PO systems, and in creating new equilibrium structures which are not able to develop in systems of colloidal self-assembly.[27] Large scale flexible photonic crystal films, potentially incorporating a wide range of engineered polymeric composite particles, can thus be made with much less stringent requirements for monodispersity.

**Experimental Section**

*Optical Microscopy*: Bright field reflection measurements are taken with a custom-modified BX51 microscope, using a 5x objective with NA ~ 0.13. The size of the spot samples an area ~ 20 μm in diameter. The microscope is coupled to an Ocean Optics QE65000 spectrometer via a QP50 UV-VIS fibre, and the reflection spectra are normalised using a silver mirror as reference. Measurements are performed with free standing PO films without backing foils. The measured reflectance is the normal reflection from the surface of the samples with a range of ±7.5°.

*X-ray Scattering*: Small angle x-ray scattering (SAXS) measurements are done on the DUBBLE beamline BM26B at the European Synchrotron Radiation Facility (ESRF), using a photon energy of 12.4 keV. Due to problems of the low scattering contrast of the PO film, no compound refractive lenses are used in the experiment, but careful beam focusing at the detector position is performed. The size of the beam is ~ 300 μm in diameter and the detector is placed approximately 7m from the sample. A Pilatus 1M 2D CCD detector, with pixel size 22 μm$^2$, is used to collect the diffraction data. PO alloys with different degrees of structural order are characterized, with samples mounted on a custom polycarbonate sample holder with an 8 mm diameter aperture in the middle, allowing the beam to penetrate the sample without touching the sample holder. Samples are scanned in the (x,y) plane with a single orientation (normal incidence) to ensure film uniformity. In order to determine the 3D order of the samples, two methods are used. High resolution SAXS measurements are performed with different sample orientations for different samples at a pre-selected (x,y) position. Also, measurements are performed by aligning the orientation of the surface of the samples to be parallel with the incident beam, in order to determine the intensity distribution in the stacking direction of the planes in reciprocal space. Further measurements are done with selected samples stained with $RuO_4$ for better contrast. The exposure time for each measurement is 60 seconds, with background measurements taken every 30 minutes by measuring the scattering in air without samples. Diffraction data is calibrated and then processed with appropriate background subtraction, and the normalized intensity of each scattering pattern obtained. Relative errors up to 2.5% in the calibration arise due to variation of the intensity of the beam between background measurements as well as from the instruments. Finally, diffracted spots are fitted by using a 2D Gaussian model.

*Focused ion beam - Scanning electron microscopy*: Samples for focused ion beam - scanning electron microscope (FIB-SEM) characterization are prepared by the following procedures. Small strips of PO films are cut from different samples and fully exposed to $RuO_4$ vapour in a sealed container for 48 hours. The stained samples are set in wax and trimmed by cryo-microtome until cross sections of the samples are exposed and a relative smooth surface is obtained. The stubs are then stained with RuO4 vapour for a further 24 hours, in order to ensure full staining of the exposed cross section surface. The stubs are then coated with a thin layer of platinum and further trimmed by using an FEI 200 FIB milling system. Milling currents from 200 pA to 800 pA under 30 kV high voltage are used for coarse and fine milling, in order to obtain a clean cross sectional surface. The stubs are then stained for a further 12 hours and coated with an ultra-thin layer of platinum for FIB-SEM (also called dual-beam or cross-beam) characterization. The FIB-SEM characterization uses a Zeiss Cross-Beam system with 50 pA to 200 pA milling current; SEM images are obtained under a voltage of 3 kV. Samples are sliced at intervals of ~20 nm by FIB, and imaged with the SEM detector at the same time. Subsequent reconstruction of the 3D images is completed using ImageJ software.

**Supporting Information**
Supporting Information is available from the Wiley Online Library or from the author.

**Acknowledgements**
The authors thank Dr Pierre Burdet for assistance with the FEI Helios NanoLab SEM-FIB system, and Dr David Snoswell for many discussions. We acknowledge grants EPSRC EP/G060649/1, EP/L027151/1, EP/G037221/1, and ERC grant LINASS 320503.

Received: ((will be filled in by the editorial staff))
Revised: ((will be filled in by the editorial staff))
Published online: ((will be filled in by the editorial staff))

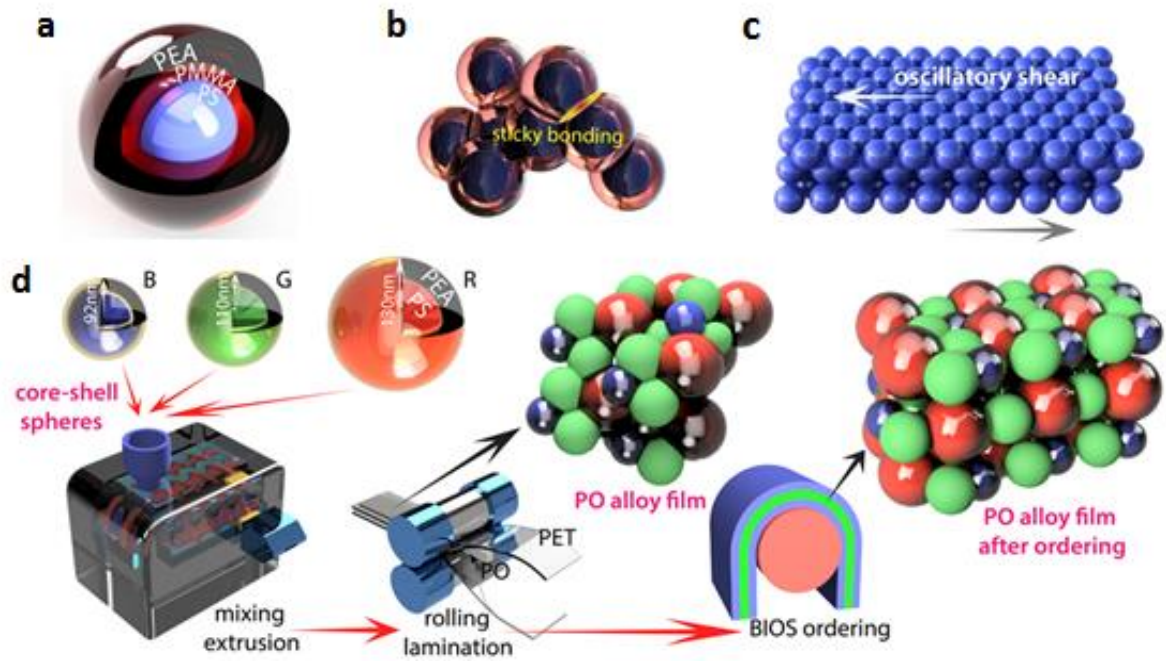

**Figure 1.**
(a) Schematic core-interlayer-shell (CIS) particle of polystyrene/poly (methylmethacrylate)/ polyethylacrylate (PS-PMMA-PEA), which have a short-range 'sticky bonding' interactions as shown in (b). (c) Illustration of mechanism of shear-ordering via oscillatory shear process. (d) Fabrication process of polymer opal alloys, showing the extrusion, rolling-lamination, and BIOS stages of sample production. B, G, and R represent core-shell spheres of different sizes that make blue, green, and red single component POs, respectively.

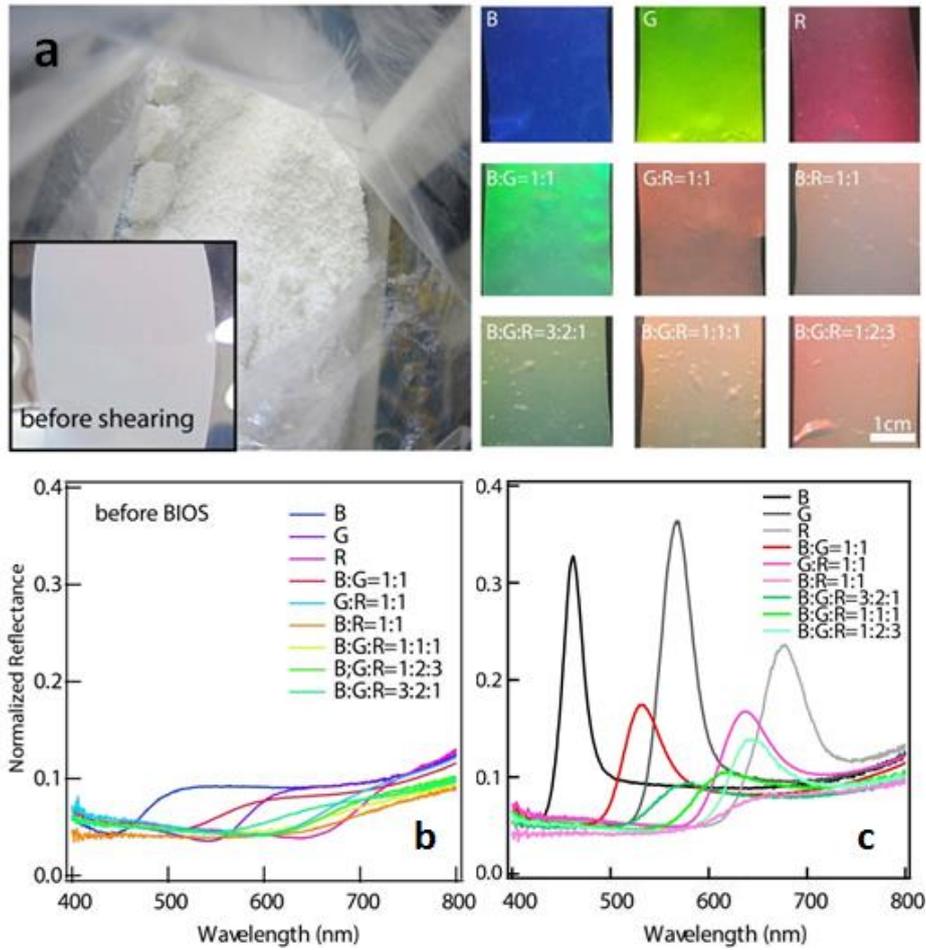

**Figure 2.**
(a) Digital photos of PO precursors and resultant rolled film before the BIOS ordering process (left), and digital photos of BIOS-processed POs with different Blue (B), Green (G) and Red (R) components (right). Binary and tertiary alloys are mixed with different component number ratios, as indicated. (b) Normalized normal-incidence reflectance spectra of PO alloys and single component POs, as indicated, before BIOS. (c) Corresponding spectra after BIOS.

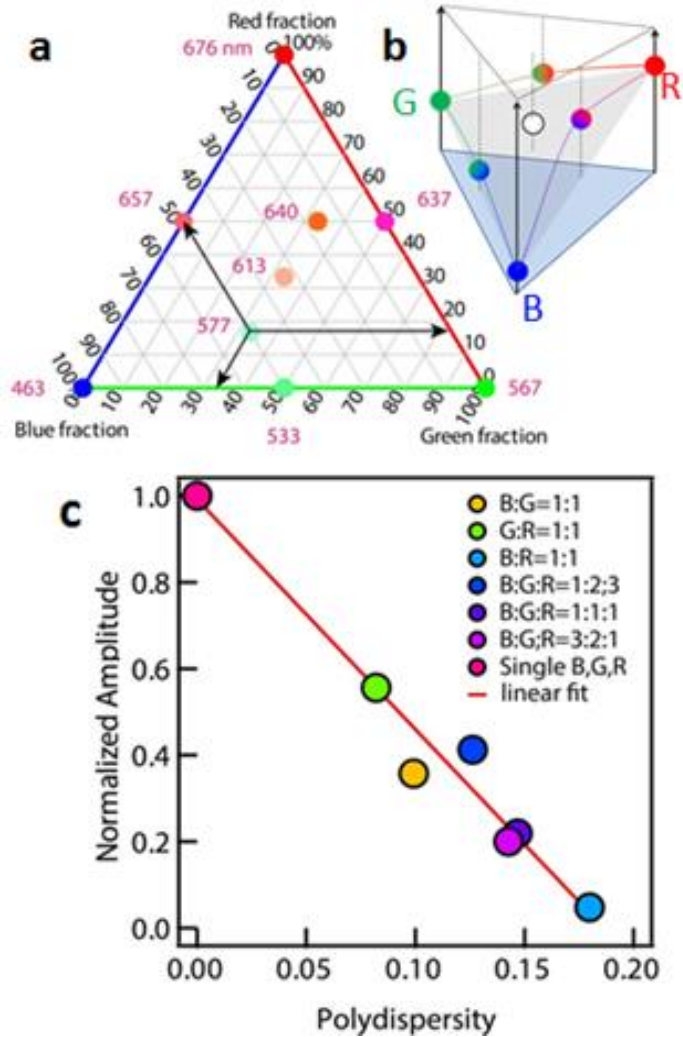

**Figure 3.**
(a) Ternary diagram of the peak wavelength of PO alloys with different components; the BGR alloys in the ratios 3:2:1, 1:1:1 and 1:2:3 have peaks at 577, 613, and 640 nm respectively. (b) Illustrates the same information as a 3D 'color space' chart, showing clear bowing along each of the 3 edges (RG, GB, and BR). (c) Normalized Bragg peak amplitude vs. sample polydispersity; the colored points denote the various binary and ternary alloys, as per the figure legend.

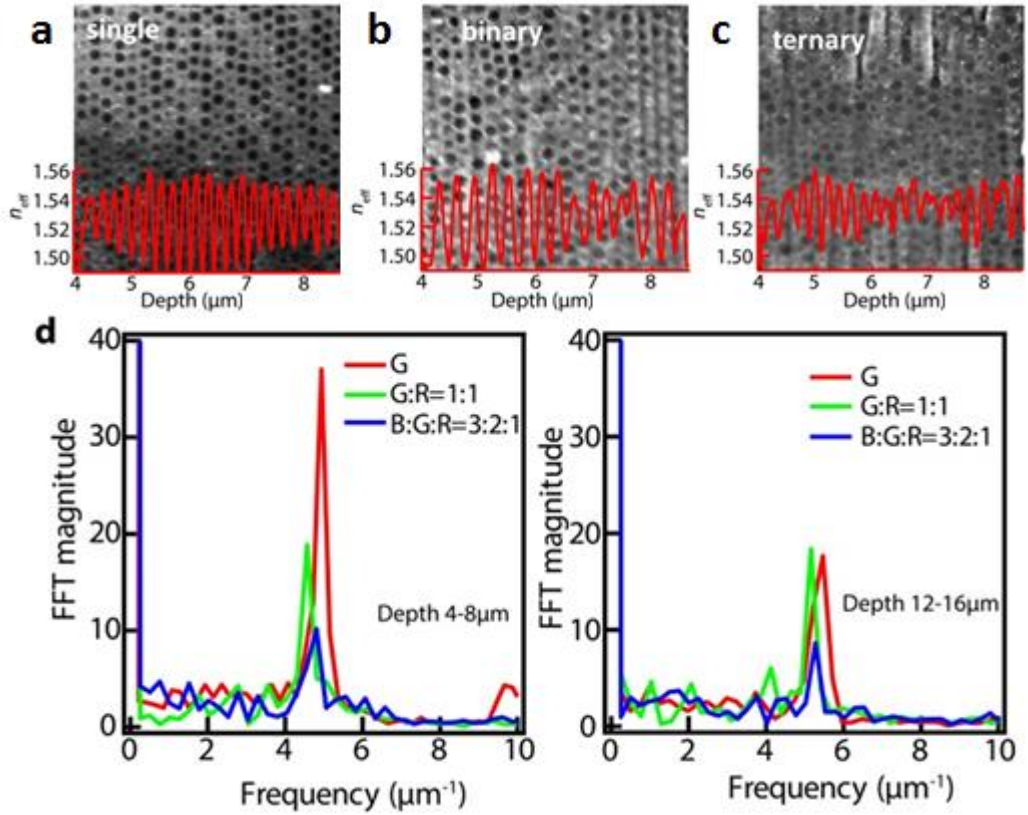

**Figure 4.**
FIB-SEM characterization of the cross sections of single component, binary, and ternary PO alloys. Cross sections are shown in the depth direction from 4-8μm from (a) a green single-component PO, (b) a binary PO alloy with G:R = 1:1, and (c) a ternary PO alloy with B:G:R = 3:2:1. Overlaid plots (in red) show effective refractive index ($n_{eff}$) along the depth direction. (d) Fast Fourier transforms (FFT) of the $n_{eff}(z)$ profile in the depth direction for different samples.

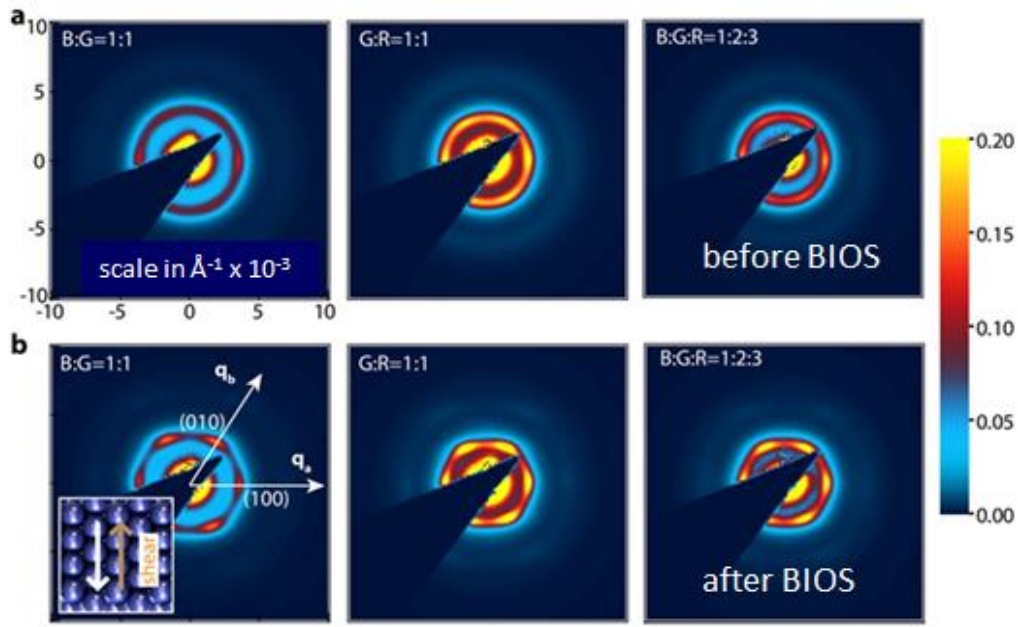

**Figure 5.**
SAXS intensity patterns of binary and ternary PO alloys (a) before and (b) after the BIOS process, at normal incidence. Intensities are adjusted to the same arbitrary range (inset right), although no further normalization between images is applied. Inset in (b) indicates the direction of oscillatory shearing relative to real space crystal orientation, with the most closely packed (mcp) direction of the spheres being along the shearing direction.

**Ordered binary and ternary photonic crystals,** composed of different sized polymer-composite spheres with diameter ratios up to 120%, are generated using bending induced oscillatory shearing (BIOS). This viscoelastic system creates polydisperse equilibrium structures, producing mixed opaline colored films with greatly reduced requirements for particle monodispersity, and very different sphere size ratios, compared to other methods of nano-assembly.

**Keyword** Photonic Crystals

Qibin Zhao, Chris E. Finlayson, Christian Schafer, Peter Spahn, Markus Gallei, Lars Herrmann, Andrei Petukhov, and Jeremy J. Baumberg*

**Nanoassembly of Polydisperse Photonic Crystals based on Binary and Ternary Polymer Opal Alloys**

ToC figure

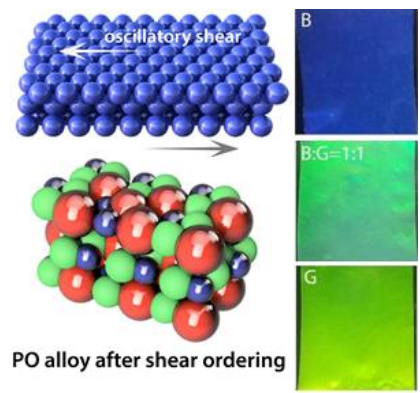

## Supporting Information

for *Adv. Opt. Mater.*, DOI: 10.1002/adom.((please add manuscript number))

**Nanoassembly of Polydisperse Photonic Crystals based on Binary and Ternary Polymer Opal Alloys**

*Qibin Zhao, Chris E. Finlayson, Christian Schafer, Peter Spahn, Markus Gallei, Lars Herrmann, Andrei Petukhov, and Jeremy J. Baumberg\**

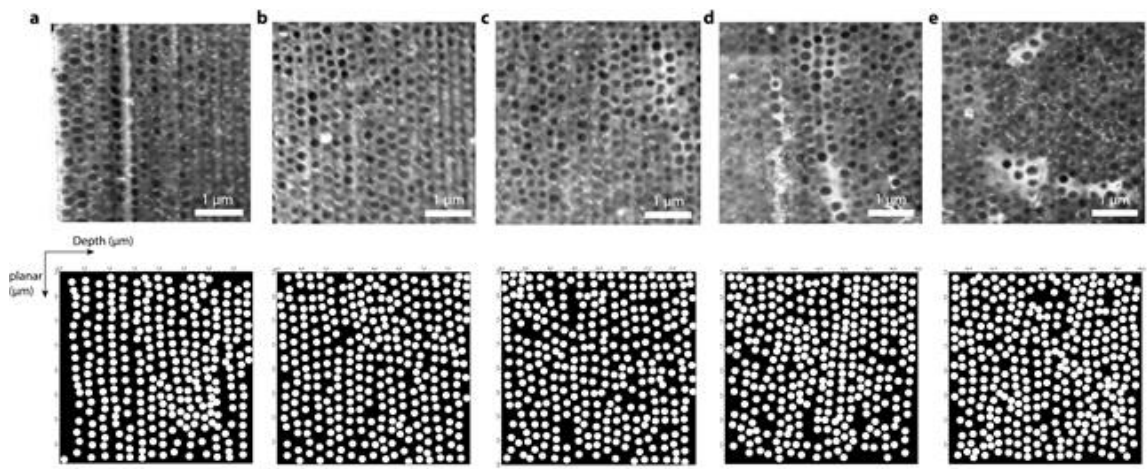

**Figure S1.** SEM images of the cross sections of G:R = 1:1 PO alloy sample with 40 passes of BIOS, cross sections trimmed by FIB. (a – e) Cross-sections in depth direction from 0 μm to 20 μm, each image being 4μm deep. Binary images below each SEM image show positions of spheres after fitting, images at same scale as SEM images, sphere diameters set to 182 nm.

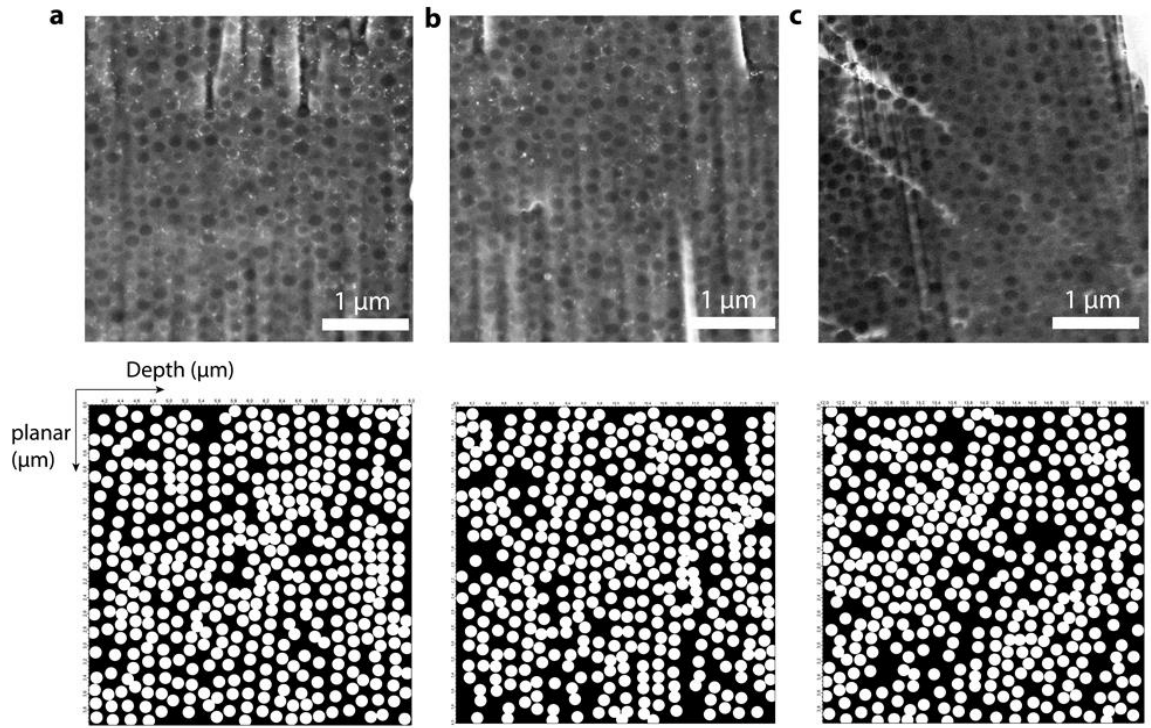

**Figure S2**. SEM images of cross sections of B:G:R = 3:2:1 PO alloy sample with 40 passes of BIOS, cross sections trimmed by FIB. (a – c) Cross-section in the depth direction from 4 μm to 16 μm, each image being 4μm deep. Binary images below each SEM image show positions of spheres after fitting, images at same scale as SEM images, sphere diameters set to 182 nm

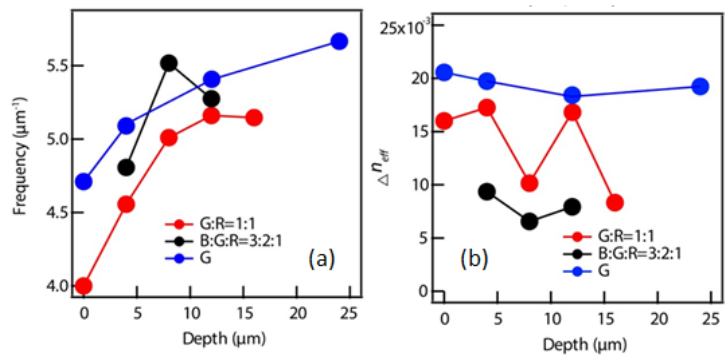

**Figure S3**. (a) Extracted spatial frequency in depth direction for single, binary and ternary POs. (b) Extracted effective refractive index contrast $\Delta n_{eff}$ along depth direction of single, binary and ternary POs.

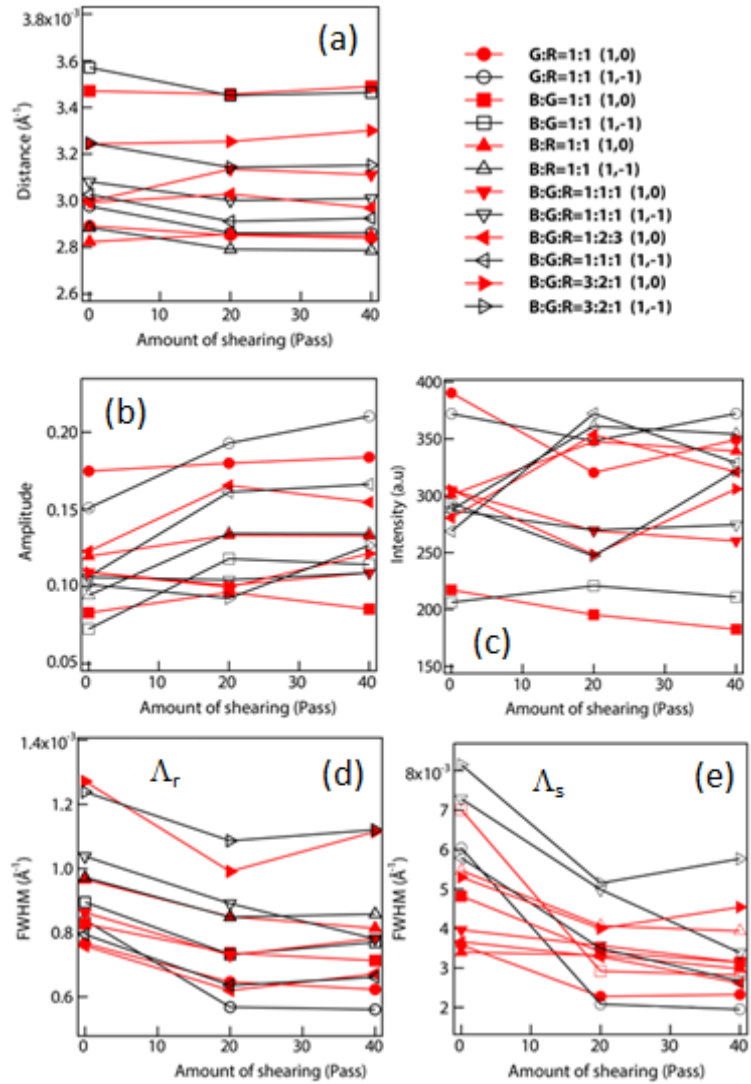

**Figure S4**. Analysis of SAXS results for binary and ternary PO alloys before and after BIOS at normal incidence. (a) Change of distance from centres of spots (10) and (1$\bar{1}$) to origin in reciprocal space with increasing BIOS shearing. (b) Change of spot amplitudes with increasing BIOS passes. (c) Change of spot intensities (areas) with increasing BIOS passes. (d) Change of FWHM in radial direction of spots $\Lambda_r$ (short axis of spots) with increasing BIOS shearing. (e) Change of FWHM in long axis (azimuthal direction) of spots $\Lambda_s$ with increasing BIOS passes.

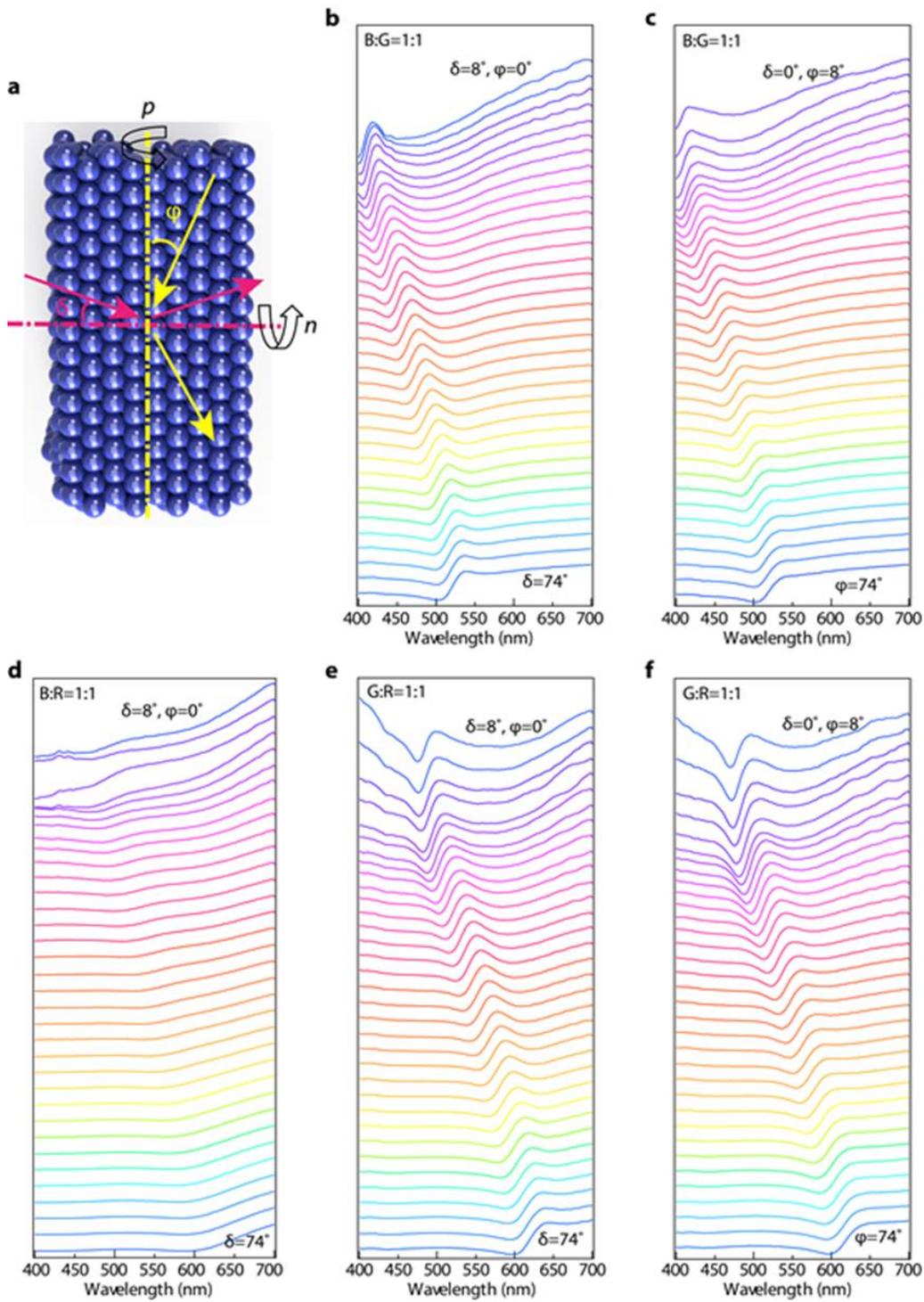

**Figure S5**. Angle-dependent specular reflection of different binary PO alloys measured with goniometer. (a) Schematic view of measurement geometry, (b-f) measured spectra of different samples at different orientations, as indicated.